\documentclass{aa}

\usepackage{txfonts}
\usepackage[dvips]{graphicx}
\usepackage{natbib}
\bibpunct{(}{)}{;}{a}{}{,} %following the A&A style.

\begin{document}

\title{The mid IR -- hard X-ray correlation in AGN and its implications for
       dusty torus models
       \thanks{Based on ESO observing programmes 075.B-0844(C) and 
               077.B-0137(A)}}
\author{Hannes Horst \inst{1,2,3}
  \and Poshak Gandhi \inst{4}
  \and Alain Smette \inst{3}
  \and Wolfgang J. Duschl \inst{1,5}}
\authorrunning{H. Horst et al.}
\offprints{H. Horst, \email{hhorst@astrophysik.uni-kiel.de}}
\institute{Institut f\"ur Theoretische Physik und Astrophysik, 
       Christian-Albrechts-Universit\"at zu Kiel, Leibnizstr. 15, 24098 Kiel,
       Germany
  \and Zentrum f\"ur Astronomie, ITA, Universit\"at Heidelberg, 
       Albert-Ueberle-Str. 2, 69120 Heidelberg, Germany
  \and European Southern Observatory, Casilla 19001, Santiago 19, Chile
  \and RIKEN Cosmic Radiation Lab, 2-1 Hirosawa, Wakoshi Saitama 351-0198,
       Japan
  \and Steward Observatory, The University of Arizona, 933 N. Cherry Ave, 
       Tucson, AZ 85721, USA}
\date{Received 00.00.0000 / Accepted 00.00.0000}
\abstract{}
         {We investigate mid-infrared and X-ray properties of the dusty torus 
          invoked in the unification scenario for active galactic nuclei by
          using the relation between mid IR and hard X-ray luminosities to 
          constrain the geometry and physical state of the dusty torus.}
         {We present new VISIR observations of 17 nearby AGN and combined 
          these with our earlier VISIR sample of 8 Seyfert galaxies. After 
          combining these observations with X-ray data from the literature, we 
          studied the correlation between their mid IR and hard X-ray 
          luminosities.}
         {A statistically highly significant correlation is found between the 
          rest frame $12.3 \, \mu$m ($L_{\mathrm{MIR}}$) and 2-10 keV 
          ($L_{\mathrm{X}}$) luminosities. Furthermore, with a
          probability of 97 \%, we find that Sy 1 and Sy 2 nuclei have the same
          distribution of $L_{\mathrm{MIR}}$ over $L_{\mathrm{X}}$.}
         {The high resolution of our MIR imaging allows us to exclude
          any significant non-torus contribution to the AGN mid IR continuum,
          thereby implying that the similarity in the 
          $L_{\mathrm{MIR}}$ / $L_{\mathrm{X}}$ ratio between Sy 1s and
          Sy 2s is intrinsic to AGN. We argue that this is best explained by 
          clumpy torus models. The slope of the correlation is in good 
          agreement with the expectations from the unified scenario and 
          indicates little to no change in the torus geometry with luminosity. 
          In addition, we demonstrate that the high angular resolution is 
          crucial for AGN studies in the IR regime.}

\keywords{galaxies: active -- Infrared: galaxies -- X-rays: galaxies}

\maketitle

\section{Introduction}

The unification model for active galactic nuclei (AGN) interprets the 
different appearance of Seyfert 1 (Sy1) and Seyfert 2 (Sy2) galaxies uniquely 
as the result of an orientation effect \citep{anton93,barthel94,urry96}. The 
central engine is considered to be surrounded by an optically and geometrically
thick molecular torus. Associated with this torus are large masses of dust 
that supposedly reprocess the X-ray and UV emission from the accretion disk 
and re-emit it in the mid infrared (MIR) regime \citep{pier93}. 

It is thus very attractive to search for correlations between IR continuum and 
hard X-ray emission in order to test the unification scenario for AGN. A tight
correlation between the $10.5 \, \mu$m continuum and the absorption-corrected 
2-10 keV luminosities for 8 nearby Seyfert galaxies was reported by 
\citet{krabbe01} using $1\farcs 2$ resolution MIR imaging. More recently, 
\citet{lutz04} found a correlation between the rest frame $6 \, \mu$m 
luminosity and the absorption-corrected hard X-ray luminosity for a sample 
of $71$ AGN. This sample was comprised of objects for which $24 \arcsec$ 
angular resolution ISOPHOT spectra and hard X-ray observations were available; 
in particular it does not contain Compton-thick objects. However, the authors
reported two problems that the unification scenario faces when compared  with 
their observations. (I) The scatter of the relation is about an order of 
magnitude larger than expected from the results of \citet{krabbe01}. (II) 
There is no significant difference between type 1 Seyferts (Sy types 1 to 1.5) 
and type 2 Seyferts (Sy types 1.8 to 2) objects in the average ratio of 
mid-infrared to hard X-ray emission, as would be expected from an optically
and geometrically thick as well as smooth torus dominating the mid IR AGN 
continuum. This was also supported by \citet{alonso02} who investigated the 
correlation between the $10 \, \mu$m luminosity and the black hole mass of
AGN. In an earlier paper \citep[][hereafter Paper~I]{horst06}, 
\defcitealias{horst06}{Paper~I} we presented VISIR observations of 8 nearby 
Seyfert nuclei. With our angular resolution of $0 \farcs 35$ at FWHM we then 
improved upon \citet{krabbe01} and \citet{lutz04} by a factor of 3 and 80, 
respectively, thus minimising contributions from extra-nuclear emission. We 
found a strong correlation between the rest frame $12.3 \, \mu$m and 2-10 keV 
luminosities with type 1 and type 2 Seyferts following the same correlation of 
$L_{\mathrm{MIR}} \propto L_{\mathrm{X}}^{1.60 \pm 0.22}$. We interpreted this 
as a strong indication for the obscuring medium to be clumpy rather than 
smoothly distributed and thus appearing as optically thin in the mid infrared. 
Moreover, we found the slope to be in good agreement with the theoretical 
predictions by \citet{beckert04}.

Here we present results from our enlarged AGN sample which contains 21 newly 
observed objects, out of which 17 were detected with VISIR. Thus, we can now 
analyse the mid IR and hard X-ray properties of 25 objects. 

Throughout this paper we assume $H_{0} = 73$ km s$^{-1}$ Mpc$^{-1}$, 
$\Omega_{\Lambda} = 0.72$ and $\Omega_{\mathrm{m}} = 0.24$ \citep{spergel06}.

\section{Target selection and X-ray data} \label{sample}

\subsection{Target selection}  \label{selection}

Our first sample of VISIR targets \citepalias[see][]{horst06} was drawn from 
the sample by \citet{lutz04}. Our criteria were observability at low airmass 
from Paranal observatory during ESO period 75, redshifts below 0.1 and 
coverage of a wide span of hard X-ray luminosities in both type 1 and type 2 
objects. For the present work we have revisited the X-ray properties of these 
objects (see Appendix \ref{indxrays}) to account for the most recent 
observations. However, only minor changes had to be applied to the
luminosities reported in \citetalias{horst06}. The revisited characterisation
of the sample is given in Table \ref{sample1}. Redshifts $z$ were taken from 
the NED, Seyfert types are according to \citet{veron06} -- with L denoting 
LINER type AGN -- and the inclination angles $i$ of the host galaxies on the 
sky were taken from the HyperLeda database \citep{paturel03}. The last column 
contains the linear scale on the plane of the sky that is resolved by VISIR, 
computed for an angular resolution of $0.35 \arcsec$, which is a typical value 
for our observations (see Table \ref{data}).

For our second sample of objects (listed in Table \ref{sample2}), we 
also selected targets not included in the Lutz et al. sample. We browsed the
literature for relatively nearby AGN ($z \leq 0.1$) with inferred 
absorption-corrected hard X-ray luminosities that are observable at low airmass
from Paranal observatory during ESO period 77. An additional criterion was the
inclination of the host galaxy toward the plane of the sky. We set the limit
at an inclination angle of $i = 65^{\circ}$ in order to avoid viewing the AGN 
through large amounts of gas in the host galaxy. Note, that two objects from 
our first sample -- \object{NGC 526a} and \object{NGC 7314} -- do not meet 
this requirement. 

\begin{table}
  \begin{center}
    \caption{Revisited characterisation of our galaxy sample for ESO period 
             75. }
    \begin{tabular}{lccccc}
      \hline\hline
      Object & $z$ & Sy & $i\,^*$ & log $L_{2-10 \, \textrm{\tiny{keV}}}$ & 
      Scale$^{**}$\\
       & & & & [erg/s] & [pc] \\

      \hline
      \object{Fairall 9} & 0.047 & 1.2 & $55.7^{\circ}$ &
      $43.87 \pm 0.15$ & 320 \\
      \object{NGC 526a} & 0.019 & 1.9 & $74.0^{\circ}$ &
      $43.14 \pm 0.10$ & 135 \\
      \object{NGC 3783} & 0.010 & 1.5 & $29.9^{\circ}$ &
      $43.21 \pm 0.15$ & 70 \\
      \object{NGC 4579} & 0.005 & L & $39.0^{\circ}$ &
      $41.10 \pm 0.15$ & 35 \\
      \object{NGC 4593} & 0.009 & 1.0 & $36.0^{\circ}$ &
      $42.93 \pm 0.20$ & 65 \\
      \object{PKS 2048-57} & 0.011 & 1h & $48.4^{\circ}$ &
      $42.84 \pm 0.20$ & 80 \\
      \object{PG 2130+099} & 0.062 & 1.5 & $63.0^{\circ}$ & 
      $43.65 \pm 0.20$ & 415\\
      \object{NGC 7314} & 0.005 & 1h & $70.3^{\circ}$ &
      $42.20 \pm 0.15$ & 35 \\
      \hline
      \multicolumn{6}{l}{ \rule[0mm]{0mm}{5mm} $^*$Column 4 lists the 
       inclination angles $i$ of the host galaxies on the sky.} \\
      \multicolumn{6}{l}{ $^{**}$Column 6 contains the scale we resolve in our
              observations.}
    \end{tabular}
    \label{sample1}
  \end{center}
\end{table}

For our target selection we searched the literature for AGN with inferred
2-10 keV luminosities. Observations at different epochs were taken as an
indicator for the intrinsic variability of the object. This task requires a
lot of care as X-ray observations can be of varying quality especially if
non-imaging instruments are involved. Furthermore, different authors sometimes 
pursue different strategies for fitting the observed data. In general we gave 
higher priority to recent data obtained at the best spatial resolution and
highest signal-to-noise ratio. Where multi-epoch observations are 
not available we assume a variability / uncertainty of a factor of 2 (0.3 dex) 
which is a typical value for our sample. Furthermore we account for a distance 
uncertainty of 10 \%.

The main characteristics of the selected AGN are listed in Table 
\ref{sample2}; a more detailed description of the X-ray properties of each 
object can be found in Appendix \ref{indxrays}. The classification of all AGN
in our sample has been done according to \citet{veron06} with the exception of
\object{NGC 4303} (see section \ref{peculiar}) and \object{Cen A} which is
listed as a possible BL Lac object but is commonly referred to as an optical 
type 2 AGN. For the rest of this paper, we summarise Sy types 1.0, 1.2 and
1.5 as ``type 1'' and Sy types 1.8, 1.9, 2.0 and 1h (broad lines detected 
in the polarised spectrum) as ``type 2''.

\begin{table}
  \begin{center}
    \caption{Characterisation of our galaxy sample for ESO period 77. Columns
             are as in Table \ref{sample1}.}

    \begin{tabular}{lccccc}
      \hline\hline
      Object & $z$ & Sy & $i$ & log $L_{2-10 \, \textrm{\tiny{keV}}}$ &
      Scale \\ & & & & [erg/s] & [pc] \\
      \hline
      \object{MCG-01-01-043} & 0.030 & 1.0 & 
      $29.9^{\circ}$ & $42.50 \pm 0.15$ & 200 \\
      \object{Mrk 590} & 0.026 & 1.0 & $25.9^{\circ}$ & 
      $43.61 \pm 0.25$ & 175 \\
      \object{NGC 1097} & 0.004 & L & $51.2^{\circ}$ & 
      $40. 80 \pm 0.15$ & 30 \\
      \object{NGC 4303} & 0.005 & L$^*$ & $19.1^{\circ}$ &
      $39.16 \pm 0.15$ & 35 \\
      \object{NGC 4472} & 0.003 & 2.0 & $50.1^{\circ}$ &
      $\leq 39.17^*$ & 20 \\
      \object{NGC 4507} & 0.012 & 1h & $34.1^{\circ}$ & 
      $43.30 \pm 0.15$ & 80 \\
      \object{NGC 4698} & 0.003 & 2.0 & $58.7^{\circ}$ &
      $39.08 \pm 0.30$ & 20 \\
      \object{NGC 4941} & 0.004 & 2.0 & $53.5^{\circ}$ &
      $41.30 \pm 0.30$ & 30 \\
      \object{IRAS 13197-1627} & 0.017 & 1h & $55.0^{\circ}$ &
      $42.78 \pm 0.20$ & 115 \\
      \object{Cen A} & 0.002 & 2.0$^*$ & $49.1^{\circ}$ &
      $41.68 \pm 0.15$ & 7$^*$ \\
      \object{NGC 5135} & 0.014 & 2.0 & $44.8^{\circ}$ &
      $43.00 \pm 0.50$ & 95 \\
      \object{MCG-06-30-15} & 0.008 & 1.5 & $60.6^{\circ}$ &
      $42.57 \pm 0.20$ & 55 \\
      \object{NGC 5995} & 0.025 & 1.9 & $41.3^{\circ}$ &
      $43.54 \pm 0.15$ & 170 \\
      \object{ESO 141-G55} & 0.036 & 1.0 & $40.0^{\circ}$ &
      $43.90 \pm 0.15$ & 240 \\
      \object{Mrk 509} & 0.034 & 1.5 & $36.3^{\circ}$ &
      $44.10 \pm 0.15$ & 225 \\
      \object{NGC 7172} & 0.009 & 2.0 & $57.8^{\circ}$ &
      $42.76 \pm 0.40$ & 60 \\
      \object{NGC 7213} & 0.006 & L & $28.6^{\circ}$ &
      $42.23 \pm 0.15$ & 40 \\
      \object{3C 445} & 0.056 & 1.5 & $36.3^{\circ}$ & 
      $44.19 \pm 0.15$ & 365 \\
      \object{NGC 7469} & 0.016 & 1.5 & $43.0^{\circ}$ &
      $43.15 \pm 0.10$ & 110 \\
      \object{NGC 7674} & 0.029 & 1h & $24.0^{\circ}$ &
      $44.56 \pm 0.50$ & 195 \\
      \object {NGC 7679} & 0.017 & 2.0 & $58.5^{\circ}$ &
      $42.52 \pm 0.15$ & 115\\
      \hline
 
      \multicolumn{6}{l}{ \rule[0mm]{0mm}{5mm} $^*$Refer to subsection 
      \ref{peculiar} for details.}
    \end{tabular}
    \label{sample2}
  \end{center}
\end{table}

\subsection{Peculiarities of individual objects} \label{peculiar}

While for most AGN the type and distance have been unambiguously determined, 
a few objects show significant discrepancies in the literature. Therefore, in 
the following, we will discuss peculiarities of some individual objects.

\object{NGC 4303}: The nucleus of this object is at the borderline between 
                being a Seyfert 2.0 AGN and a LINER galaxy 
                \citep{filippenko85}. Since it is not evident -- although very 
                likely -- that the nuclear X-ray source is an AGN, we adopt
                the classification as a LINER, in contrast to \citet{veron06}.

\object{NGC 4472}: This object was erroneously included in our sample. An
                   X-ray flux taken from the literature turned out to originate
                   in a Ultraluminous X-ray source; the nucleus has not
                   been unambigously detected in the X-rays
                   \citep[but see][]{maccarone03}.

\object{Cen A}: The redshift of Centaurus A is $z = 0.001825$. In our cosmology
             this corresponds to a luminosity distance of 7.49 Mpc. However,
             the recessional movement of the galaxy is mostly peculiar and not
             due to the Hubble flow. The best available distance estimate 
             toward Cen A is $d = 3.84$ Mpc. It was derived by 
             \citet{rejkuba04}, using two independent methods: The 
             period-luminosity relation for 
             Mira variables and the luminosity of the tip of the red giant 
             branch. The uncertainty in Cen A's distance together with -- by 
             now outdated -- different cosmological parameters have led some 
             authors to overestimate the luminosity of this object by more 
             than one order of magnitude. 
             
\object{3C 445}: This Broad Line Radio Galaxy is an interesting case: 
              \citet{sambruna98} and \citet{shinozaki06} both invoked two 
              different absorbing components to fit their respective X-ray 
              spectra. Moreover, the ratio of optical reddening to hydrogen
              column density is more than one order of magnitude beneath the
              galactic value. This indicates an anomalous gas-to-dust ratio 
              \citep{maiolino01}. The most recent X-ray observations
              by \citet{grandi07} and \citet{sambruna07}, again, indicate the
              presence of multiple absorbing layers with differing covering
              factors.
       
\object{NGC 7674}: NGC 7674 appears to be a Compton-thick AGN of high 
                luminosity (see Appendix \ref{indxrays} for details). Strictly
                speaking, it does not match our target selection criteria as
                its intrinsic hard X-ray luminosity cannot be inferred 
                with good precision. It is, however an interesting test case 
                in order to check whether the MIR -- hard X-ray correlation 
                holds for Compton-thick objects.
 
\object{NGC 7679}: Another interesting object; being of Seyfert type 1.9 (broad
                H$_{\alpha}$ line but no broad H$_{\beta}$) one would expect 
                the X-ray spectrum to show a low-energy cutoff typical for
                absorbed systems. However, in their analysis \citet{dellac01} 
                found NGC 7679 to appear as a Seyfert 1 in X-rays. Moreover,
                they found evidence for a starburst in the nucleus of this 
                object.

\section{Observations and data analysis} \label{data}

Our first sample of objects was observed between April and August 2005; 
details on observing conditions can be found in 
\citetalias{horst06}. The second sample was observed one year later, between 
April and August 2006. We used the standard imaging template of VISIR, with 
parallel chopping and nodding and a chop throw of $8\arcsec$. In order to get 
the best possible angular resolution, the small field camera ($0\farcs075$ / 
pixel) was used. Bright AGN were observed in three filters in order to allow a 
reconstruction of their spectral energy distribution (SED) in the MIR. Due to
time constraints, faint objects could only be observed in one filter. All 
observations were executed in service mode with required observing conditions
of clear sky and $0\farcs8$ seeing or less. The average airmass was 1.15, with 
no observation being executed at an airmass above 1.3. Science targets and 
photometric standards were all observed within 2 h of each other and with a 
maximum difference in airmass of 0.25. For most observations, however, 
differences in both time and airmass are much smaller than these values.

Some exposures had to be re-executed due to changing atmospheric conditions. In
these cases, for each object only the data obtained under the best conditions
were used for our subsequent analysis. A log of these observations is given in
Table \ref{obslog}.

\begin{table*}
  \caption{Basic observational parameters for all data used from our 2006 
           campaign. STD and Obj denote parameters for standard star and
           science observations, respectively. The FWHMs are the average of 
           all beams of one exposure.}
           
  \begin{center}
    \begin{tabular}{lclcccccc}
      \hline\hline
      Object & Obs. Date & Filter$^*$ & \multicolumn{2}{c}{FWHM [$\arcsec$]} &
      \multicolumn{2}{c}{Airmass} & Flux & log $L_{12.3 \mu \mathrm{m}}$ \\
      & (MM-DD) & & STD & Obj & STD & Obj & [mJy] & [erg s$^{-1}$] \\
      \hline
      \object{MCG-01-01-043} & 07-13 & NeII & 0.37 & -- & 1.014 & 1.064 & 
              $\leq 68.0$ & $\leq 44.52$ \\
      \object{Mrk 590} & 08-17 & SIV & 0.29 & 0.24 & 1.071 & 1.094 & 
              $75.9 \pm 20.9$ & -- \\
      \object{Mrk 590} & 08-17 & PAH2 & 0.30 & 0.27 & 1.073 & 1.094 & 
              $75.0 \pm 2.1$ & -- \\
      \object{Mrk 590} & 08-17 & NeII & 0.33 & 0.32 & 1.075 & 1.094 &
              $106.3 \pm 13.3$ & $43.55 \pm  0.05$ \\
      \object{NGC 1097} & 08-07 & NeIIref1 & 0.32 & 0.37 & 1.037 & 1.051 &
              $28.2 \pm 6.8$ & $41.40 \pm 0.10$ \\
      \object{NGC 4303} & 04-15 & NeIIref1 & 0.40 & -- & 1.155 & 1.147 &
              $\leq 22.0$ & $\leq 42.40$ \\
      \object{NGC 4472} & 04-18 & NeIIref1 & 0.34 & -- & 1.171 & 1.238 &
              $\leq 77.0$ & $\leq 41.62$ \\
      \object{NGC 4507} & 04-15 & SIV & 0.32 & 0.32 & 1.151 & 1.052 &
              $523.2 \pm 24.9$ &  -- \\
      \object{NGC 4507} & 04-15 & PAH2 & 0.33 & 0.32 & 1.147 & 1.049 &
              $589.5 \pm 21.8$ & -- \\
      \object{NGC 4507} & 04-15 & NeIIref1 & 0.34 & 0.34 & 1.143 & 1.047 &
              $685.0 \pm 50.1$ & $43.67 \pm 0.03$ \\
      \object{NGC 4698} & 04-18 & NeIIref1 & 0.34 & -- & 1.171 & 1.196 &
              $\leq 42.0$ & $ \leq 42.54$ \\
      \object{NGC 4941} & 04-19 & NeIIref1 & 0.34 & 0.35 & 1.045 & 1.131 &
              $81.3 \pm 6.0$ & $41.74 \pm 0.05$ \\
      \object{IRAS 13197-1627} & 04-09 & SIV & 0.29 & 0.36 & 1.012 & 1.018 &
              $527.1 \pm 17.1$ & -- \\
      \object{IRAS 13197-1627} & 04-09 & PAH2 & 0.32 & 0.38 & 1.012 & 1.021 &
              $674.3 \pm 35.8$ & -- \\
      \object{IRAS 13197-1627} & 04-09 & NeIIref1 & 0.34 & 0.43 & 1.013 &
              1.023 & $875.0 \pm 45.8$ & $44.07 \pm 0.02$ \\
      \object{Cen A} & 04-09 & SIV & 0.29 & 0.32 & 1.012 & 1.054 &
              $642.6 \pm 26.6$ & -- \\
      \object{Cen A} & 04-09 & PAH2 & 0.32 & 0.36 & 1.012 & 1.055 &
              $946.6 \pm 29.2$ & -- \\
      \object{Cen A} & 04-09 & NeIIref1 & 0.34 & 0.35 & 1.013 & 1.056 &
              $1451 \pm 73.1$ & $41.80 \pm 0.03$ \\
      \object{NGC 5135} & 04-09 & NeIIref1 & 0.34 & 0.35 & 1.013 & 1.021 &
              $122.5 \pm 12.2 $ & $43.06 \pm 0.04$ \\
      \object{MCG-06-30-15} & 04-14 & SIV & 0.30 & 0.32 & 1.110 & 1.034 &
              $339.2 \pm 43.7$ & -- \\
      \object{MCG-06-30-15} & 04-14 & PAH2 & 0.31 & 0.33 & 1.103 & 1.030 &
              $392.5 \pm 54.1$ & -- \\
      \object{MCG-06-30-15} & 04-14 & NeIIref1 & 0.32 & 0.35 &  1.096 & 1.027 &
              $392.7 \pm 49.3$ & $43.07 \pm 0.05$ \\
      \object{NGC 5995} & 04-14 & SIV & 0.30 & 0.37 & 1.110 & 1.195 &
              $296.8 \pm 30.2$ & -- \\
      \object{NGC 5995} & 04-14 & PAH2 & 0.33 & 0.37 & 1.022 & 1.181 &
              $332.9 \pm 47.2$ & -- \\
      \object{NGC 5995} & 04-14 & NeII & 0.37 & 0.40 & 1.090 & 1.169 &
              $421.1 \pm 60.6$ & $44.11 \pm 0.06$ \\
      \object{ESO 141-G55} & 05-05 & SIV & 0.37 & 0.37 & 1.043 & 1.206 &
              $160.0 \pm 21.2$ & -- \\
      \object{ESO 141-G55} & 05-05 & PAH2 & 0.37 & 0.36 & 1.045 & 1.206 &
              $169.8 \pm 23.9$ & -- \\
      \object{ESO 141-G55} & 05-05 & NeIIref1 & 0.37 & 0.32 & 1.046 & 1.207 &
              $169.7 \pm 47.1$ & $44.04 \pm 0.11$ \\
      \object{Mrk 509} & 06-14 & SIV & 0.47 & 0.32 & 1.260 & 1.253 &
              $226.5 \pm 7.7$ & -- \\
      \object{Mrk 509} & 06-14 & PAH2 & 0.39 & 0.33 & 1.282 & 1.270 &
              $235.0 \pm 21.4$ & -- \\
      \object{Mrk 509} & 06-14 & NeII & 0.40 & 0.38 & 1.318 & 1.288 &
              $269.0 \pm 41.7$ & $44.18 \pm 0.06$ \\
      \object{PKS 2048-57} & 05-05 & SIV & 0.37 & 0.38 & 1.043 & 1.199 &
              $590.6 \pm 19.4$ & -- \\
      \object{PKS 2048-57} & 05-05 & PAH2 & 0.37 & 0.38 & 1.045 & 1.196 &
              $752.4 \pm 45.5$ & -- \\
      \object{PKS 2048-57} & 05-05 & NeIIref1 & 0.37 & 0.43 & 1.046 & 1.193 &
              $1035 \pm 107.9$ & $43.82 \pm 0.04$ \\
      \object{NGC 7172} & 07-09 & NeIIref1 & 0.37 & 0.35 & 1.187 & 1.188 &
              $164.9 \pm 27.1$ & $42.79 \pm 0.07$ \\
      \object{NGC 7213} & 07-14 & SIV & 0.35 & 0.31 & 1.031 & 1.272 &
              $283.8 \pm 6.2$ & -- \\
      \object{NGC 7213} & 07-14 & PAH2 & 0.34 & 0.35 & 1.034 & 1.284 &
              $264.0 \pm 38.5$ & -- \\
      \object{NGC 7213} & 07-14 & NeIIref & 0.32 & 0.32 & 1.040 & 1.298 &
              $271.0 \pm 26.5$ & $42.67 \pm 0.05$ \\
      \object{3C 445} & 07-10 & SIV & 0.29 & 0.31 & 1.147 & 1.129 &
              $168.4 \pm 6.7$ & -- \\
      \object{3C 445} & 07-10 & PAH2 & 0.30 & 0.32 & 1.147 & 1.135 &
              $184.6 \pm 10.4$ & -- \\
      \object{3C 445} & 07-10 & NeII & 0.33 & 0.37 & 1.148 & 1.143 &
              $205.8 \pm 27.8$ & $44.50 \pm 0.06$ \\
      \object{NGC 7469} & 07-12 & SIV & 0.31 & 0.38 & 1.324 & 1.250 &
              $460.0 \pm 20.0 $ & -- \\
      \object{NGC 7469} & 07-12 & PAH2 & 0.33 & 0.41 & 1.334 & 1.243 &
              $487.3 \pm 38.6$ & -- \\
      \object{NGC 7469} & 07-12 & NeIIref1 & 0.34 & 0.37 & 1.344 & 1.237 &
              $626.9 \pm 34.7$ & $43.92 \pm 0.02$ \\
      \object{NGC 7674} & 07-13 & NeII & 0.38 & 0.44 & 1.150 & 1.199 &
              $506.3 \pm 29.4$ & $44.31 \pm 0.02$ \\
      \object{NGC 7679} & 07-10 & SIV & 0.28 & 0.24 & 1.147 & 1.140 &
              $42.4 \pm 13.0$ & -- \\
      \object{NGC 7679} & 07-10 & PAH2 & 0.30 & 0.18 & 1.147 & 1.143 &
              $43.3 \pm 6.6$ & -- \\
      \object{NGC 7679} & 07-10 & NeIIref1 & 0.32 & 0.41 & 1.147 & 1.146 &
              $45.6 \pm 18.3$ & $42.82 \pm 0.15$ \\
      \hline

      \multicolumn{9}{l}{ \rule[0mm]{0mm}{5mm} $^*$The central wavelengths
                          for the individual filters are $9.82 \, \mu$m for 
                          SIVref1, $10.49 \, \mu$m for SIV, $11.25 \, \mu$m
                          for PAH2,} \\
      \multicolumn{9}{l}{$11.88 \, \mu$m for PAH2ref2, $12.27 \, \mu$m for 
                         NeIIref1 and $12.81 \, \mu$m for NeII. Their 
                         transmission curves are shown in the} \\
      \multicolumn{9}{l}{VISIR User Manual at 
                          http://www.eso.org/instruments/visir/.}
    \end{tabular}

  \end{center}
  \label{obslog} 
\end{table*}

We reduced science and standard star frames using the pipeline written by Eric
Pantin (private communication) for the VISIR consortium. To eliminate
glitches, the pipeline applies a bad pixel mask and removes detector stripes. 
Subsequently we removed background variations using a 2 dimensional polynomial 
fit. For objects observed in unstable conditions we treated each nodding 
cycle separately as the background pattern sometimes changed between two 
consecutive cycles. The count rate for one full exposure was calculated as the 
mean of all 3 beams from all nodding cycles of this exposure. As an error 
estimate we use the standard deviation of these. In order to minimise the 
effect of residual sky background we chose relatively small apertures 
($\approx 10$ pixels $= 1 \farcs 27$) for the photometry and corrected the 
obtained count rates using the radial profiles of standard stars. Finally we 
calibrated our photometry using the same standard stars. The conversion factor 
counts/s / Jy proved to be very stable: for each individual filter, variations 
were less than 10 \% over the whole observing period.

\section{Results} \label{results}

\subsection{Mid IR properties}  \label{basics}

Out of the 21 AGN observed with VISIR, 17 were detected. They are all 
point-like which means the torus remains unresolved. For the 4 non detected 
objects $3 \sigma$ upper limits were derived by measuring the flux of the 
largest noise-induced signal and assuming the same PSF as measured for the 
standard stars. In Table \ref{obslog} the flux or flux limit and size of the 
PSF is listed for each object. For sources observed in three filters, we were 
able to reconstruct their MIR SEDs. For a few objects we find
indications for the presence of the $9.7 \, \mu$m feature, either in emission
or in absorption. Looking at our own data and comparing with archival
\emph{Spitzer} spectra, we find that in all cases, at wavelengths longer than
$12 \, \mu$m, the feature hardly affects the observed flux level. Possible
contamination will be well below 10 \%. 

In addition to the central point source, in three sources -- 
\object{NGC 1097}, \object{NGC 5135}, \object{NGC 7469} -- we observed clear
extra-nuclear MIR emission. In all of these objects we find distinct blobs of 
star formation (SF), in addition NGC 5135 and NGC 7469 also exhibit a weak 
diffuse component. Typical distances between SF blobs and the AGN are 
$8\farcs2$ ($\sim 700$ pc) in NGC 1097, $1\farcs5$ ($\sim 400$ pc) in NGC 5135 
and $1\farcs3$ ($\sim 400$ pc) in NGC 7469. In the latter two objects the 
proximity of the SF components to the AGN does not allow to rule out a 
significant contribution of SF to the measured flux of the central point 
source. By comparing the fluxes of individual SF blobs and of the AGN, we can 
estimate upper limits for such contamination. This yields 15 \% for NGC 5135 
and 10 \% for NGC 7469. These uncertainties have been added to the photometric 
error and are accounted for in our subsequent statistical analysis. 

Details on the morphologies and SEDs of all observed objects will be presented
in a forthcoming paper. We should, however, point out that the 
chopping/nodding technique used for the observations would automatically 
cancel a contribution from a smooth and diffuse component extending over an 
area similar to the chopping/nodding throw.

\subsection{The mid IR -- hard X-ray correlation} \label{stats}

In order to analyse the correlation between intrinsic hard X-ray and mid 
infrared luminosity, for each object we have taken the photometry done closest
to our reference wavelength of $12.3 \, \mu$m in rest frame and then converted 
to said wavelength by assuming the SED to be flat in this part of the
spectrum. This wavelength was chosen as it is close to the peak of the MIR SED 
in AGN and, in addition, is only affected by emission or absorption
in one of the silicate features at $9.7 \, \mu$m and $18 \, \mu$m if these are
particularly strong. Fortunately, there is no indication for such 
strong silicate features in any observed SED.

\begin{figure}
  \begin{center}
    \includegraphics[width=0.45\textwidth]{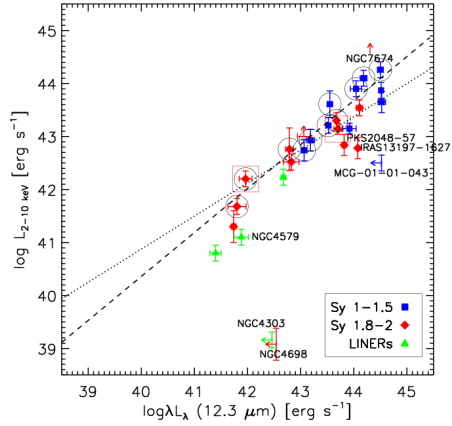}
    \caption{Correlation of MIR and absorption-corrected hard X-ray 
             luminosities for our VISIR sample. Blue squares are type 1
             Seyferts (type 1.5 or smaller), red diamonds are type 2
             Seyferts; green triangles are LINERs. The two galaxies with large
             inclination angles (see section \ref{selection}) are marked by a 
             red square. Well resolved objects (see section \ref{contamination}
             for details) are marked by a black 
             circle. These have been used for the displayed power-law fit 
             (dashed line); the dotted line is the best fit to our first 
             sample, as discussed in \citetalias{horst06}. Arrows indicate 
             either upper limits to the MIR luminosity or lower limits to the 
             X-ray luminosity; the arrows' colour code corresponds to the one 
             of the other symbols. Note that \object{NGC 4472} is not shown 
             here since neither its X-ray nor MIR luminosities could be 
             determined.}
   \label{mirxcorrel}
  \end{center}
\end{figure}

The resulting monochromatic luminosity is plotted versus the intrinsic 2-10
keV luminosity in Fig. \ref{mirxcorrel}. The correlation between these two 
quantities is of high statistical significance; after excluding non-detections
and Compton-thick objects, the Spearman rank coefficient \citep{spearman04} is $\rho = 0.90$ at a 
significance level of $7.3 \times 10^{-9}$. We can, therefore, safely reject
the null hypothesis of no intrinsic correlation. The best power-law fit to the 
whole sample, excluding \object{NGC 526a} and \object{NGC 7314} because of the 
large inclination angles of their host galaxies and thus consisting of 10 
type 1s, 11 type 2s and 4 LINERs, is $\log L_{\mathrm{MIR}} = 
(-1.61 \mp 1.85) + (1.04 \pm 0.04) \log L_{\mathrm{X}}$. For all objects not 
detected with VISIR, the $3 \sigma$ flux limits are compatible with this 
correlation, including the two low luminosity objects \object{NGC
4303} and \object{NGC 4698}.

It is important to note, that while the extended sample further strengthens
the statistical significance of the correlation we had found in our first
sample \citepalias{horst06}, the slope of the correlation has changed
considerably. After excluding NGC 4579 due to its peculiar nature, our first 
sample yielded $L_{\mathrm{MIR}} \propto L_{\mathrm{X}}^{1.60 \pm 0.22}$ which 
is consistent with our new slope at the 3$\sigma$ level. The discrepancy in 
slope is caused by our first sample being small and containing some objects
(i.e. \object{PKS 2048-57}) which are off the correlation we find for the 
enlarged sample. The latter agrees well with the results of the earlier
studies by \citet{krabbe01} and \citet{lutz04}. 

One concern with correlations between luminosities is that they may be
caused by a redshift bias in the data. We tested for this possibility in two
ways. First of all, we analysed the formal correlation between MIR and hard 
X-ray fluxes. The resulting Spearman rank coefficient is $\rho = 0.52$ at a 
significance level of 0.01. This means that the correlation is rather weak but,
nevertheless, highly significant. Secondly, we tried to reproduce the observed
correlation between luminosities with randomly generated mock data. For this,
we assumed X-ray and MIR flux values to be uniformly distributed over the
range of observed ones, for the distances we tried both a uniform and a normal
distribution with no significant change in the result: After generating 
$10^{4}$ mock datasets, a rank coefficient of $\rho > 0.89$ for the MIR -- hard
X-ray correlation was found with a frequency of less than 0.1 \%, a
significance level of less than $7.4 \times 10^{-9}$ with a frequency of 
less than 0.15 \%. The combination of both results -- the high significance of 
the flux -- flux correlation and the low probability of reproducing the
obserevd luminosity -- luminosity correlation with non-correlated data --
implies that we can safely assume the correlation to be real and not caused by
a redshift bias in our data.

\begin{figure}
  \begin{center}
    \includegraphics[width=0.35\textwidth,angle=90]{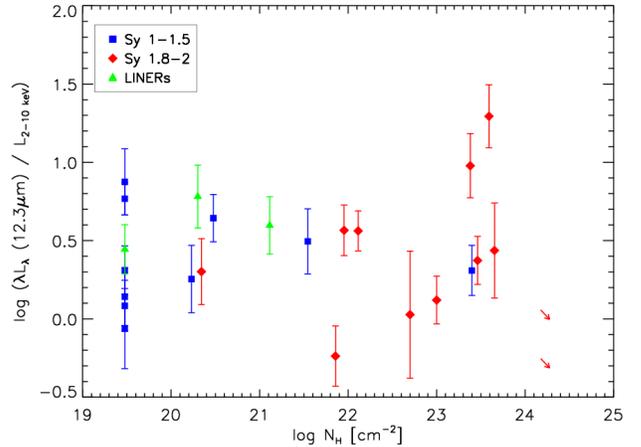}
    \caption{Luminosity ratio $L_{\mathrm{MIR}} /  L_{\mathrm{X}} \, vs$  
             intrinsic column density $N_{H}$. Symbols and colours are as in 
             Fig. \ref{mirxcorrel}. Compton-thick objects are displayed as red 
             arrows, non-detections in the infrared have been omitted. They
             would be positioned outside of the displayed region, toward higher
             luminosity ratios. Objects with no measured intrinsic absorption 
             have been placed at $N_H = 10^{19.5}$ cm$^{-2}$ for clarity.}
    \label{mirxnh}
  \end{center}
\end{figure}

In Fig. \ref{mirxnh} we show the luminosity ratio $L_{\mathrm{MIR}} /
L_{\mathrm{X}} \, vs$ the intrinsic column density $N_H$ toward the AGN. There 
is no correlation discernible between the two displayed quantities. Neither do 
we find a significant dependence of the luminosity ratio on Seyfert type:
$\left< \log L_{\mathrm{MIR}} - \log L_{\mathrm{X}}\right>$ is 
$\left( 0.38 \pm 0.31 \right)$ for type 1s, $\left( 0.44 \pm 0.45 \right)$
for type 2s and $\left( 0.61 \pm 0.17 \right)$ for the three detected LINERs
in our sample. We tested for a possible dependence of luminosity
ratio on column density by computing the Spearman rank coefficient for
it. With a result of $\rho = 0.13$ at a significance level of 0.54, we find no
evidence for a correlation between the two quantities. Furthermore, we
performed a Kolmogorov-Smirnov test for the luminosity ratios of type 1 and
type 2 Seyferts. The maximum difference in the cumulative distribution
function is 0.2 and the significance level 0.97 which means both samples very
likely originate from the same parent distribution. These numbers confirm the 
result from our first study that type 1s and type 2s follow the same 
$L_{\mathrm{MIR}} - L_{\mathrm{X}}$ correlation and have the same average 
luminosity ratio.

\section{Discussion} \label{discussion}

\subsection{Contamination} \label{contamination}

A potential problem for our study is contamination by extra-nuclear emission.
This is primarily important in the mid infrared. AGN of moderate to high 
luminosities should always dominate their hosts in the 2-10 keV band. In the 
mid infrared, on the other hand, star formation and also Narrow Line Region 
(NLR) clouds can  significantly contribute to the total flux.

First of all, the high angular resolution (see Tables \ref{sample1} and
\ref{sample2} for the physical scales we resolve) cuts away most of the star
formation. The off-nuclear emission we find in three of our objects (see 
section \ref{basics}) is at distances from the AGN that are resolved even in
the most distant objects of our sample. Therefore, we are confident that
emission from SF regions does not heavily affect our measurements. 

\begin{figure}
  \begin{center}
    \includegraphics[width=0.35\textwidth,angle=90]{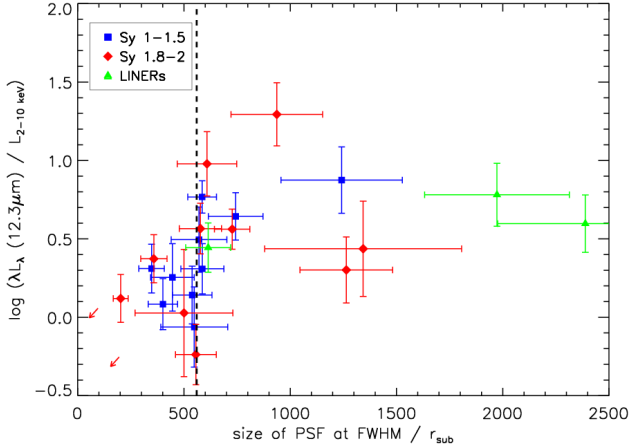}
    \caption{Luminosity ratio $L_{\mathrm{MIR}} /  L_{\mathrm{X}} \, vs$
             resolved physical scale in units of $r_{\mathrm{sub}}$. Symbols 
             and colours are as in Fig. \ref{mirxcorrel}. As in Fig. 
             \ref{mirxnh}, Compton-thick objects are displayed as red arrows
             and MIR  non-detections have been omitted. The vertical dashed
             line indicates a resolution of $560 \cdot r_{\mathrm{sub}}$
             above which the luminosity ratio rises.}
    \label{mirxscale}
  \end{center}
\end{figure}

The NLR, however, remains unresolved in most of our sources. The case of NGC 
1068 shows that this may be a problem, as very accurate MIR 
photometry of this object \citep{galliano05} show the NLR clouds to contribute
almost 50 \% of the nuclear MIR emission. Fortunately, three of our detected 
sources, namely \object{NGC 4579}, \object{NGC 1097} and, most importantly,
\object{Cen A}, are at comparable or even smaller distances than NGC 1068 and
do not show any additional nuclear components. In order to test whether less
well resolved sources are affected by extra-nuclear emission, in Fig. 
\ref{mirxscale} we have plotted the MIR / X-ray luminosity ratio over the 
resolved physical scale, expressed in units of the dust sublimation radius 
$r_{\mathrm{sub}} \propto L_{\mathrm{bol}}^{1/2}$. We here assume the 
accretion disc's bolometric luminosity to be $L_{\mathrm{bol}} = 10 \cdot 
L_{\mathrm{X}}$ which is a typical value for Seyfert galaxies
\citep{vasudevan07}. Interestingly, for observations for which the size of the
PSF at FWHM (in pc) is less than $560 \cdot r_{\mathrm{sub}}$, there is no
case with $L_{\mathrm{MIR}} / L_{\mathrm{X}} > 0.4$, while this is the case
for all but two objects with FWHM (pc) $> 560 r_{\mathrm{sub}}$. The 
mean luminosity ratios for the well resolved and the less well resolved 
sources are $\left( 0.11 \pm 0.19 \right)$ and $\left( 0.65 \pm 0.27 \right)$, 
respectively. This rise indicates that some of our less well resolved
sources are, indeed, affected by contamination, despite our high angular 
resolution. 

We have, therefore, recalculated the mid IR -- hard X-ray correlation, only 
using well resolved objects with FWHM (pc) / $r_{\mathrm{sub}} \leq
560$ (encircled objects in Fig. \ref{mirxcorrel}). The resulting
$\log L_{\mathrm{MIR}} = (-3.89 \mp 3.68) + (1.09 \pm 0.09) 
\log L_{\mathrm{X}}$ is in very good agreement with the correlation found for 
the whole sample. This method allows us to say with some confidence that 
objects situated close to this fit (dashed line in Fig. \ref{mirxcorrel}) are 
not affected by contamination. Interestingly, for the well resolved
sources we find the correlation between the measured fluxes to be much
stronger than for the complete sample. Its Spearman rank is $\rho = 0.92$ and 
the significance level $ = 5 \times 10^{-4}$. This is another indication that 
these fluxes are mostly free of contamination.

The SF region we find in three of our objects and also Fig. \ref{mirxscale} 
illustrate the need for high angular resolution MIR studies for testing AGN 
models. Even using 8m-class telescopes, a good sample selection is crucial to 
avoid contamination by non-AGN emission. Furthermore, this indicates that the 
high scatter in the mid IR -- hard X-ray correlation found by \citet{lutz04} 
may, indeed, be caused by the poor angular resolution of their 
ISOPHOT data. 

Despite the difference in slope we have found in 
\citetalias{horst06}, we again find \object{NGC 4579} -- the object we had 
excluded from the power-law fit in our first study -- to lie off the 
correlation (see Fig. \ref{mirxcorrel}).

\subsection{Is the torus clumpy or smooth?} \label{clumps}

There has been a long-going debate over whether the matter in the obscuring
tori in AGN is distributed smoothly \citep{pier92} or arranged in clumps 
\citep{krolik88}. In \citetalias{horst06} we argued that type 1 and type 2 
nuclei having the same $L_{\mathrm{MIR}} / L_{\mathrm{X}}$ ratio, is a strong
argument for the torus to be clumpy. This point is strengthened by our current
study not only due to the improved number statistics but also due to the
larger range of column densities we probe. As discussed in section
\ref{stats}, the ratio $L_{\mathrm{MIR}} / L_{\mathrm{X}}$ shows no dependence
on $N_H$ -- and so over 4 orders of magnitude in $N_H$ (see 
Fig. \ref{mirxnh}). Even the two Compton-thick objects in our sample, 
\object{NGC 5135} and \object{NGC 7674} seem to have roughly the same value of 
$L_{\mathrm{MIR}} / L_{\mathrm{X}}$ as the rest of our sample. For these 
objects, however, strong conclusions cannot be drawn, as their X-ray 
luminosities can only be estimated to an uncertainty of one order or magnitude.

For an optically thick smooth torus, a much higher 
$L_{\mathrm{MIR}} / L_{\mathrm{X}}$ is expected for Sy 1s than for Sy 2s. 
\citet{pier93} expect a difference of one order of magnitude for a
  change of $N_{\mathrm{H}}$ from $\sim 10^{20}$ cm$^{-2}$ to $\sim 10^{24}$ 
cm$^{-2}$. The reason for this is that in Sy 2s the observer sees the emission
from the cold dust in the outer part of the torus, while in Sy 1s one can see 
its hot inner part as well. This prediction is not compatible with our 
results.

It is important to note that the similarity between Sy types still holds if we 
only regard objects with 
FWHM (pc) / $r_{\mathrm{sub}} \leq 560$. For these we find 
$\left< \log L_{\mathrm{MIR}} - \log L_{\mathrm{X}} \right>$
to be $\left( 0.15 \pm 0.15\right)$ for type 1s and 
$\left( 0.07 \pm 0.25 \right)$ for type 2s.

Models of clumpy tori \citep{nenkova02,dullemond05,hoenig06} do not predict
a difference in $L_{\mathrm{MIR}} / L_{\mathrm{X}}$ between different Sy
types. In particular, \citet{hoenig06}, using 3D radiative transfer modelling, 
showed that clumpy tori can appear as optically thin in the MIR with most of 
the radiation originating in the innermost part of the torus. In their model 
individual clouds are optically thick but their volume filling factor is
small. Our observational results clearly prefer this kind of models to smooth 
ones.

Recently, \citet{ramos07} reported to have found a difference in $\left< 
\log L_{6.75 \mu \mathrm{m}} - \log L_{\mathrm{X}} \right>$ between type 1 and 
type 2 objects. For their analysis, they use ISOCAM data first published by 
\citet{clavel00} and the X-ray data that was compiled by
\citet{lutz04}. Despite the brightness profile decomposition \citet{ramos07}
performed their data may still be heavily contaminated by nuclear star 
formation. Among our own data, in the two cases of \object{NGC 5135} and 
\object{NGC 7469} we find that within $3 \arcsec$ from the nucleus, SF 
contributes at least 43 \% and 45 \% of the total flux at $12.3 \, \mu$m, 
respectively. Also the fact that the luminosity ratios \citet{ramos07} derive 
are $\sim 8$ larger than what we find for our well resolved objects -- and 
despite their shorter observing wavelength in the IR -- indicates that they
are probably affected by contamination and the MIR fluxes they report are
likely not dominated by torus emission.

\subsection{The slope of the correlation and the shape of the torus}
\label{slope}

As the mid IR radiation is accretion disc emission reproduced by the dusty 
torus, $L_{\mathrm{MIR}}$ and $L_{\mathrm{X}}$ are directly linked to each 
other via the covering factor $f_C$ of the torus:

\begin{eqnarray}
  L_{\mathrm{MIR}} \propto f_C L_{\mathrm{X}} \label{lxlmir} \mathrm{ .}
\end{eqnarray}

Thus, our best-fit result implies 
$f_{C} \propto L_{\mathrm{X}}^{0.04 \pm 0.04}$. This means that we do not find
any dependence of $L_{\mathrm{MIR}} / L_{\mathrm{X}}$ on $L_{\mathrm{X}}$
which is also illustrated by Fig. 4. This result does not match the 
expectations from the receding torus model \citep{lawrence91} for which 
\citet{simpson05} found the fraction of type 2 AGN $f_2$ to depend on AGN 
luminosity as $f_2 \propto L_{\mathrm{X}}^{-0.27}$ for $L_{\mathrm{X}} \gtrsim 
L_0 \approx 7 \times 10^{43}$ erg s$^{-1}$. The basic assumption of the 
unified scenario is that $f_2 \approx f_C$. The dependence derived by Simpson 
would, thus, yield a correlation $f_{C} \propto L_{\mathrm{X}}^{-0.27}$ which 
disagrees with the slope we find by more than $7 \sigma$. Note, however, that 
we do not probe the Quasar regime of luminosities, i.e. for a break luminosity 
$L_0 \gtrsim 10^{45}$ erg s$^{-1}$ the receding torus model would, again,
agree with our results. 

One possibility to reconcile our findings with the obvious dependence of the 
type 1 fraction on AGN luminosities \citep[][and references therein]{simpson05}
would be that a significant number of Compton-thick AGN at moderate 
luminosities ($L_{\mathrm{2-10 keV}} = 10^{43} \sim 10^{44}$ erg s$^{-1}$)
have been missed in hitherto existing X-ray surveys. The existence of 
a population of Compton-thick AGN has also been suggested by X-ray 
surveys \citep[e.g. the COSMOS survey,][]{hasinger07}, X-ray background 
synthesis \citep[e.g.][]{gilli07,gandhi07} and IR surveys \citep{martinez05}. 
Very recently, \citet{ueda07} randomly selected two AGN out of those first 
detected by \emph{Swift} to be observed with \emph{Suzaku} and found both of 
them to be highly obscured. This is anoter indication for the existence of a
large population of Compton-thick AGN at moderate luminosities.

\begin{figure}
  \begin{center}
    \includegraphics[width=0.35\textwidth,angle=90]{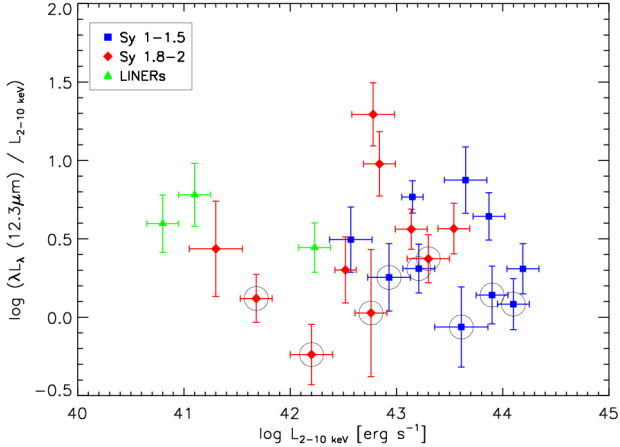}
    \caption{Luminosity ratio $L_{\mathrm{MIR}} / L_{\mathrm{X}}$, plotted
             over $L_{\mathrm{X}}$. Colours and symbols are as in Fig. 
             \ref{mirxcorrel}. Black circles mark well resolved sources with 
             FWHM(pc) / $r_{\mathrm{sub}} \leq 560$ (see section
             \ref{contamination} and Fig. \ref{mirxscale}). }
    \label{mirxcolor}
  \end{center}
\end{figure}

\subsection{Tori at low luminosities?} \label{lowl}

In paper I we found some indications that a different physical mechanism, other
than re-radiation of the accretion disk emission, is dominating the MIR 
emission in AGN with $L_{\mathrm{2-10 keV}} < 10^{42}$ erg s$^{-1}$. This 
matches theoretical predictions that have been published subsequently. 
Interestingly, the very different models by \citet{elitzur06} and 
\citet{hoenig07} both predict the collapse of the torus below 
$L_{\mathrm{bol}} \approx 10^{42}$  erg s$^{-1}$. In our new sample, however, 
the indications for this effect are less pronounced. We do see a tendency for
objects with low X-ray luminosities to depart from the MIR -- hard X-ray 
correlation, but this without statistical significance. Thus, while in 
principle, our approach is well suited to test these predictions, our current 
sample is not sufficient for this task. Future observations are required to 
test down to what luminosities the classic picture of AGN can be applied.

\section{Conclusions} \label{conclusion}

We have presented new high resolution mid infrared photometry of 17 AGN, 
increasing our total sample to 25 detected and 4 non-detected objects. We have 
found angular resolution to be a crucial requirement for our study as poorly 
resolved sources can be heavily contaminated by non-AGN emission in the mid IR 
band.

The rest frame $12.3 \, \mu$m luminosities $L_{\mathrm{MIR}}$ of our well
resolved sources strongly correlate with their rest frame 2-10 keV
luminosities $L_{\mathrm{X}}$, matching the expectations from the unified 
scenario for AGN. With a probability of 97 \%, type 1 and type 2 
Seyferts have the same distribution of 
$L_{\mathrm{MIR}} / L_{\mathrm{X}}$. This similarity is most probably 
intrinsic to the AGN and not caused by extra-nuclear emission contaminating
the MIR flux. These results have two important implications:
\begin{itemize}
  \item As we do not find the offset in $L_{\mathrm{MIR}} / L_{\mathrm{X}}$ 
        between type 1s and type 2s, predicted by smooth torus models, AGN
        tori are likely to be clumpy. The volume filling factor of these
        clumps has to be small in order to reproduce our results.
  \item At a $7 \sigma$ level, the slope of the correlation is not 
        compatible with the predictions of the receding torus model by 
        \citet{simpson05}. We suggest that the break luminosity for this model
        is higher than hitherto assumed. A population of Compton thick AGN at 
        moderate luminosities being missed in current X-ray surveys could 
        resolve this problem.
\end{itemize}
The question, whether AGN at low luminosities, with $L_{\mathrm{bol}} < 
10^{42}$ erg s$^{-1}$ have a qualitatively different appearance than objects
of higher luminosities, remains open. Fortunately, we have recently been 
awarded time to observe a sample of 14 heavily obscured 
($N_{\mathrm{H}} = 10^{24} \sim 10^{25}$ cm$^{-2}$), mostly low luminosity AGN 
with VISIR which may help to shed light on this issue.

\begin{acknowledgements}

We thank Dr. Sebastian H\"onig and Dr. Thomas Beckert for inspiring discussions
on the topic of AGN tori.
We also thank an anonymous referee for very helpful comments and 
suggestions.
Furthermore, we thank Olivier Garcet for finding and pointing out two typos in
our calculations.
H.H. acknowledges support from DFG through SFB 439. 
P.G. is a Fellow of the Japan Society for the Promotion of Science (JSPS).
This research made use of the NASA/IPAC Extragalactic Database (NED) which 
is operated by the Jet Propulsion Laboratory, California Institute of 
Technology, under contract with the National Aeronautics and Space 
Administration.
This research has made use of the Tartarus (Version 3.2) database, created by 
Paul O'Neill and Kirpal Nandra at Imperial College London, and Jane Turner at 
NASA/GSFC. Tartarus is supported by funding from PPARC, and NASA grants 
NAG5-7385 and NAG5-7067.
We acknowledge the usage of the HyperLeda database 
(http://leda.univ-lyon1.fr).
\end{acknowledgements}

\bibliographystyle{aa}
\bibliography{mybiblio}

\begin{appendix}
\section{X-ray properties of individual objects} \label{indxrays}

\subsection{X-ray data for the P75 sample}   \label{xrayp75}

\object{Fairall 9}: With XMM-\emph{Newton}, \citet{gondoin01} measured a hard 
                 X-ray flux of $F_{\mathrm{2-10 keV}} = (1.56 \pm 0.33) 
                 \times 10^{-11}$ erg s$^{-1}$ cm$^{-2}$. From the compilation
                 of earlier measurements shown in their paper, we can estimate 
                 the peak-to-peak variability / uncertainty to be of the 
                 order of 2. \citet{yaqoob04} report a flux of 
                 $F_{\mathrm{2-10 keV}} = 2.2 \times 10^{-11}$ erg s$^{-1}$ 
                 cm$^{-2}$., obtained with \emph{Chandra}. Within errors, this 
                 is still consistent with Gondoin et al.'s measurement. From 
                 these fluxes and the reported hydrogen column density of 
                 $N_{\mathrm{H}} = (3.0 \pm 0.2) \times 10^{20}$ cm$^{-2}$,
                 we calculate a luminosity of $L_{\mathrm{2-10 keV}} = 
                 7.43 \times 10^{43}$ erg s$^{-1}$.
       
\object{NGC 526A}: For this object, unfortunately, no observations with 
                \emph{Chandra} or XMM-\emph{Newton} have been published. Using 
                \emph{Beppo}SAX, \citet{landi01} measured 
                $F_{\mathrm{2-10 keV}} = 1.8 \times 10^{-11}$ erg s$^{-1}$ 
                cm$^{-2}$, absorbed by $N_{\mathrm{H}} = (1.33 \pm 0.15) \times
                10^{22}$ cm$^{-2}$. Another observation of NGC 526A with the 
                RXTE satellite has been performed by \citet{revnivtsev04}.  
                Extrapolating their 2-9 keV count rate yields 
                $F_{\mathrm{2-10 keV}} = 2.7 \times 10^{-11}$ erg s$^{-1}$ 
                cm$^{-2}$. Both observations are consistent with each other if 
                we assume an uncertainty of 0.2 dex. The resulting luminosity 
                is $L_{\mathrm{2-10 keV}} = 1.37 \times 10^{43}$ erg s$^{-1}$. 
                
\object{NGC 3783}: This object was observed with XMM-\emph{Newton} by 
                \citet{blustin02} who report an intrinsic flux of 
                $F_{\mathrm{2-10 kev}} = 8.5 \times 10^{-11}$ erg 
                s$^{-1}$ cm$^{-2}$ which corresponds to
                $L_{\mathrm{2-10 kev}} = 1.61 \times 10^{43}$ erg
                s$^{-1}$. The galactic absorption toward NGC 3783 is 
                $N_{\mathrm{H}} = 8.7 \times 10^{20}$ cm$^{-2}$. The intrinsic
                warm absorption can be ignored for our purpose as it hardly 
                affects the 2-10 keV band. Variability of this source is 
                indicated by comparing the flux given above to the one reported
                by \citet{malizia97} ($F_{\mathrm{2-10 keV}}^{\mathrm{obs}} = 
                4.09 \times 10^{-10}$ erg s$^{-1}$) which is about five times
                higher. Due to the higher quality of their data, we will use 
                the results from \citet{blustin02} and allow for a variability 
                / uncertainty of 0.3 dex.
   
\object{NGC 4579}: This object has been observed multiple times 
                \citep{terashima98, ho01, dewangan04, cappi06} with most of
                the results being in good agreement with each other. We here 
                adopt the results of Cappi et al. -- $L_{\mathrm{2-10 keV}} = 
                1.26 \times 10^{41}$ erg s$^{-1}$, $N_{\mathrm{H}} \leq 2 
                \times 10^{20}$ cm$^{-2}$ -- which were obtained with 
                XMM-\emph{Newton}. The luminosity stated by \citet{dewangan04}
                is lower by factor of 10. This, however, seems to be a
                typographical error as all measured fluxes are consistent.
                A statistical uncertainty of 0.3 dex is assumed to account for 
                the different results on $L_{\mathrm{2-10 keV}}$ among the
                other authors.
                
\object{NGC 4593}: The two most reliable data sets for this object are from 
                \citet{reynolds04} and \citet{shinozaki06}, both obtained with
                XMM-\emph{Newton}. They agree within errors. We adopt the 
                values provided by Shinozaki et al.: 
                $N_{\mathrm{H}} = 1.69 \times 10^{20}$ cm$^{-2}$ and 
                $L_{\mathrm{2-10 keV}} = 8.60 \times 10^{42}$ erg s$^{-1}$. 
                \citet{steenbrugge03} observed NGC 4593 with both 
                XMM-\emph{Newton} and \emph{Chandra}. While the \emph{Chandra} 
                result ($L_{\mathrm{2-10 keV}} = 8.91 \times 10^{42}$ erg 
                s$^{-1}$) is in good agreement with the one cited above, the 
                luminosity measured with XMM-\emph{Newton} was higher 
                ($L_{\mathrm{2-10 keV}} = 1.2 \times 10^{43}$ erg s$^{-1}$). 
                This may be due to intrinsic variability of the source. We can 
                account for this with an uncertainty of 0.4 dex.

\object{PKS 2048-57}: \citet{risaliti02b}, observing with ASCA, inferred 
                   $N_{\mathrm{H}} = (2.37 \pm 0.20) \times 10^{23}$ cm$^{-2}$ 
                   and $F_{\mathrm{2-10 keV}} = 2.65 \times 10^{-11}$ erg 
                   s$^{-1}$ cm$^{-2}$, the latter yielding 
                   $L_{\mathrm{2-10 keV}} = 6.88 \times 10^{42}$ erg s$^{-1}$.
                   An independent analysis of archival ASCA data by 
                   \citet{heckman05} yields $L_{\mathrm{2-10 keV}} = 2.88
                   \times 10^{42}$ erg s$^{-1}$, after conversion to our 
                   Cosmology. These results are consistent if we consider  
                   variability of the source: \citet{georga01} report a flux
                   variation by a factor of almost two within one week of 
                   observation. For our study we use the results by Risaliti
                   et al., allowing for a luminosity uncertainty of 0.4 dex.
           
\object{PG 2130+099}: \citet{lawson97} find this object to be unabsorbed and 
                  emit $F_{\mathrm{2-10 keV}} = 5.3 \times 10^{-12}$ erg 
                  s$^{-1}$ cm$^{-2}$ and thus $L_{\mathrm{2-10 keV}} = 4.50 
                  \times 10^{43}$ erg s$^{-1}$. More recently, \citet{gallo06b}
                  observed PG 2130+099 with XMM-\emph{Newton} and obtained 
                  $F_{\mathrm{2.5-10}} = 0.31 \times 10^{-11}$ erg s$^{-1}$ 
                  cm$^{-2}$. Using the luminosity value from Lawson \& 
                  Turner and assume an uncertainty of 0.4 dex, this is in 
                  good agreement with Gallo's result.

\object{NGC 7314}: This object was observed with XMM-\emph{Newton} by 
                \citet{shinozaki06}. After converting their result to our 
                Cosmology, we get $L_{\mathrm{2-10 keV}} = 1.5 \times 10^{42}$ 
                erg s$^{-1}$ and a column density of 
                $N_{\mathrm{H}} = (7.2 \pm 1) \times 10^{21}$ cm$^{-2}$. Older 
                observations by 
                \citet{malizia97,turner97,risaliti02a,risaliti02b} and 
                \citet{kraemer04} are in good agreement with these numbers if 
                one accounts for the different cosmological parameters used.

\subsection{X-ray data for the P77 sample} \label{xrayp77}

\object{MCG-01-01-43}: This object was observed with ASCA by \citet{turner97}. 
                    They report $N_{\mathrm{H}} = 3.27^{+2.74}_{-0.0} \times 
                    10^{20}$ cm$^{-2}$ and a luminosity that -- corrected for 
                    the different Cosmologies used -- translates to 
                    $L_{\mathrm{2-10 keV}} = 3.5 \times 10^{42}$ erg s$^{-1}$.
                    This measurement is backed up by INTEGRAL observations  
                    \citep{ebisawa03} where JEM-X measures 
                    $F_{\mathrm{3-10 keV}} = 1.55 \times 10^{12}$ erg s$^{-1}$ 
                    cm$^{-2}$. This yields $L_{\mathrm{3-10 keV}} = 2.84 
                    \times 10^{42}$ erg s$^{-1}$.

\object{Mrk 590}: With XMM-\emph{Newton}, \citet{gallo06} estimated a 2-10 keV
               luminosity which, in our Cosmology, yields 
               $L_{\mathrm{2-10 keV}} = 6.6 \times 10^{42}$ erg s$^{-1}$. This
               is in good agreement with the results of a combined 
	       \emph{Chandra} and XMM-\emph{Newton} programme by 
               \citet{longinotti07} ($L_{\mathrm{2-10 keV}} = 8.9 \times 
               10^{42}$ erg s$^{-1}$). \citet{shinozaki06}, on the other hand, 
               having observed Mrk 590 with XMM-\emph{Newton} as well, report 
               a luminosity of $L_{\mathrm{2-10 keV}} = 4.1 \times 10^{43}$ 
               erg s$^{-1}$ (after conversion to our Cosmology). Here, we will 
               use Longinnotti et al.'s result and allow for an uncertainty of 
               0.5 dex.
  
\object{NGC 1097}: \citet{iyomoto96} observed this object with ASCA and 
                retrieved $N_{\mathrm{H}} = 1.3^{+0.4}_{-0.3} \times 10^{21}$ 
                cm$^{-2}$ and $F_{\mathrm{2-10 keV}} = 1.7 \times 10^{-12}$
                erg s$^{-1}$ cm$^{-2}$, the latter corresponding to 
                $L_{\mathrm{2-10 keV}} = 6.18 \times 10^{40}$ erg s$^{-1}$. 
                \citet{nemmen06} observed NGC 1097 with \emph{Chandra} and 
                report $N_{\mathrm{H}} = 2.3^{+2.8}_{-1.7} \times 10^{20}$ 
                cm$^{-2}$ and $F_{\mathrm{2-10 keV}} = 1.73 \times 10^{-12}$ 
                erg s$^{-1}$ cm$^{-2}$. We will adopt these results for our 
                analysis. 
 
\object{NGC 4303}: This low-luminosity AGN was observed by \citet{jimenez03}, 
                using the \emph{Chandra} telescope. The nuclear source does
                not show intrinsic absorption, thus the galactic value of
                $N_{\mathrm{H}} = 1.67 \times 10^{20}$ cm$^{-2}$ is
                adopted. The unabsorbed flux is $F_{\mathrm{2-10 keV}} = 
		2.6^{+1.0}_{-0.8} \times 10^{-14}$ erg s$^{-1}$ cm$^{-2}$. This
                corresponds to a luminosity of $L_{\mathrm{2-10 keV}} = 
                1.44 \times 10^{39}$ erg s$^{-1}$. Note, that 
                \citet{jimenez03} cannot exclude the possibility that the 
		nuclear X-ray source of NGC 4303 is a binary system instead of
                an AGN. They come, however, to the conclusion that this is 
                unlikely.

\object{NGC 4472}: This object, unfortunately, has not been detected in the 
                hard X-ray band so far. The most stringent upper limit to its
                flux is from \citet{panessa06}: $F_{\mathrm{2-10 keV}} \leq 
                6.6 \times 10^{-14}$ erg s$^{-1}$ cm$^{-2}$. The corresponding 
                luminosity limit is $L_{\mathrm{2-10 keV}} \leq 1.48 \times
                10^{39}$ erg s$^{-1}$. This is in agreement with the $3\sigma$ 
                detection in the soft band by \citet{soldatenkov03} who 
                measured $L_{\mathrm{0.5-2.5 keV}} = 1.7 \times 10^{38}$ erg 
                s$^{-1}$. Whether this source is dominated by a star-burst or 
                suffers from Compton-thick absorption is still under 
                investigation.
    
\object{NGC 4507}: For this object the estimates for the hydrogen column 
                density vary significantly. An average value is the one 
                estimated by \citet{bassani99}: $N_{\mathrm{H}} = (2.92 \pm 
                0.23) \times 10^{23}$ cm$^{-2}$. The resulting 
                absorption-corrected flux is $F_{\mathrm{2-10 keV}} = 7.03 
                \times 10^{-11}$ erg s$^{-1}$ cm$^{-2}$. This yields 
                $L_{\mathrm{2-10 keV}} = 2.0 \times 10^{43}$ erg s$^{-1}$; we 
                assume an uncertainty of 0.3 dex -- this also matches the 
                variability observed by \citet{georga01}.

\object{NGC 4698}: Three \emph{Chandra} observations of this object have been
                published, by \citet{cappi06, gonzalez06} and 
                \citet{panessa06}. Cappi et al. report $N_{\mathrm{H}} \leq
                4 \times 10^{21}$ cm$^{-2}$ and $L_{\mathrm{2-10 keV}} = 
                1.59 \times 10^{39}$ erg s$^{-1}$. If we use the distance to
                NGC 4698 from \citet{tully88} which was also used by Cappi et 
                al. the flux measurements by Panessa et al. and 
                Gonz\'{a}lez-Mart\'{i}n et al. yield 
                $L_{\mathrm{2-10 keV}} = 1.54 \times 10^{39}$ erg 
                s$^{-1}$ and $L_{\mathrm{2-10 keV}} = 4.90 \times 10^{38}$ erg
                s$^{-1}$, respectively. As all three observations seem to be
                of comparable quality we adopt a mean luminosity of 
                $L_{\mathrm{2-10 keV}} = 1.21 \times 10^{39}$ erg s$^{-1}$ and 
                an uncertainty of 0.6 dex.

\object{NGC 4941}: \citet{maiolino98} present \emph{Beppo}SAX observations of
                this object. Their best fit results are $N_{\mathrm{H}} = 
                4.5^{+2.5}_{-1.4} \times 10^{23}$ cm$^{-2}$ and
                $L_{\mathrm{2-10 keV}} \approx 2 \times 10^{41}$ erg s$^{-1}$.
                \citet{risaliti02a} complements the \emph{Beppo}SAX data with
                ASCA observations. From these he draws the conclusion that the
                absorbing column is actually Compton-thick. \citet{terashima02}
                also find $N_{\mathrm{H}} \sim 10^{24}$ cm$^{-2}$ and attribute
                the difference between their and Maiolino et al.'s result to
                variability of the absorbing column. We therefore consider the 
                uncertainty of Maiolino et al.'s result to be large and assume 
                0.6 dex for the luminosity. 

\object{IRAS 13197-1627}: The most recent X-ray observation, using
                       XMM-\emph{Newton} is by \citet{miniutti07}. They report
                       a hydrogen column density of $N_{\mathrm{H}} = (3.9 \pm
                       0.4) \times 10^{23}$ cm$^{-2}$. If we convert the
                       absorption-corrected luminosity they infer to our 
                       Cosmology we get
                       $L_{\mathrm{2-10 keV}} = (0.6 \pm 0.2) \times 10^{43}$ 
                       erg s$^{-1}$. Miniutti et al. also estimate the true
                       intrinsic luminosity of the AGN by multiplying the 
                       reflection fraction. After correcting for the different
                       Cosmologies it is $L_{\mathrm{2-10 keV}}^{\mathrm{int}}
                       = (3.9 \pm 2.6) \times 10^{43}$ erg s$^{-1}$. The first 
                       value is also consistent with $L_{\mathrm{2-10 keV}} = 
                       0.37 \times 10^{43}$ erg s$^{-1}$ as resulting from the
                       flux reported by \citet{bassani99} -- even though these
                       authors assumed a much higher column density of 
                       $N_{\mathrm{H}} = (7.6 \pm 1.3) \times 10^{23}$ 
                       cm$^{-2}$. For our further analysis we will adopt 
                       Miniutti et al.'s absorption-corrected luminosity
                       $L_{\mathrm{2-10 keV}}$ and take the conservative 
                       assumption of an uncertainty of 0.4 dex.

\object{Cen A}: Observations with \emph{Chandra} and XMM-\emph{Newton} by 
             \citet{evans04} yield an absorbing column of $N_{\mathrm{H}} 
             \approx 10^{23}$ cm$^{-2}$ and an intrinsic luminosity of 
             $L_{\mathrm{2-10 keV}} = 4.80 \times 10^{41}$ erg s$^{-1}$ with a 
             peak-to-peak variability by a factor of $\lesssim 2$. This is 
             consistent with older estimates by \citet{grandi03} 
             ($F_{\mathrm{2-10 kev}} = 3.8 \times 10^{-10}$ erg s$^{-1}$ 
             cm$^{-2}$, yielding $L_{\mathrm{2-10 keV}} = 6.71 \times 10^{41}$ 
             erg s$^{-1}$) and \citet{risaliti02a} ($N_{\mathrm{H}} \approx 9 
             \times 10^{22}$ cm$^{-2}$). We adopt the luminosity from Evans et 
             al. and the absorbing column density from Risaliti.
             
\object{NGC 5135}: \citet{levenson04} observed NGC 5135 with \emph{Chandra}. 
                They find $N_{\mathrm{H}} > 10^{24}$ cm$^{-2}$. From the flux 
                of the Iron K$\alpha$ line they estimate the intrinsic hard 
                X-ray luminosity to be $L_{\mathrm{2-10 keV}} \approx 1 \times
                10^{43}$ erg s$^{-1}$. To account for the high uncertainty of 
                the luminosity estimate, we will set it to 1.0 dex.
                                
\object{MCG-06-30-15}: This source has been extensively studied due to its 
                    prominent, relativistically-broadened Iron K$\alpha$ line 
                    \citep[e.g.][]{tanaka89,fabian02}. A dusty warm absorber 
                    with an equivalent neutral column density of a few 
                    $\times 10^{21}$ cm$^{-2}$ is known to be present 
                    \citep{reynolds97b,lee01}. Extrapolating the latest 
                    3--45 keV Suzaku X-ray data \citep{miniutti07b} to 
                    2--10 keV results in an intrinsic power-law luminosity of 
                    $L_{\mathrm{2-10 keV}}=3.7 \times 10^{42}$ erg s$^{-1}$. 
                    The source is known to vary significantly; we have thus 
                    used time-averaged measurements for the above calculation, 
                    and also assign a variability / uncertainty of 0.4 dex. 
                    This is also consistent with an INTEGRAL/JEM-X measurement 
                    by \citet{beckmann06}.

\object{NGC 5995}: This object was observed by \citet{panessa02}, using the 
                ASCA satellite. They find $N_{\mathrm{H}} = 9.0^{+0.5}_{-0.3} 
                \times 10^{21}$ cm$^{-2}$ and (after conversion to our 
                Cosmology) $L_{\mathrm{2-10 keV}} = 3.48 \times 10^{43}$ erg 
                s$^{-1}$.

\object{ESO 141-G55}: This object was observed with XMM-\emph{Newton} by 
                   \citet{gondoin03}. They find the absorption to be Galactic 
                   with $N_{\mathrm{H}} = 5.5 \times 10^{20}$ cm$^{-2}$. 
                   After we correct the luminosity they determine for the 
                   slightly different redshift and cosmological parameters we 
                   use, we end up with $L_{\mathrm{2-10 keV}} = 8.01 \times 
                   10^{43}$ erg s$^{-1}$. 
  
\object{Mrk 509}: \citet{shinozaki06} observed this AGN with XMM-\emph{Newton} 
               and find an intrinsic absorption of $N_{\mathrm{H}} < 4.8 
               \times 10^{20}$ cm$^{-2}$ and (after correcting for slightly 
               different Cosmology and redshift) $L_{\mathrm{2-10 keV}} = 1.3
               \times 10^{44}$ erg s$^{-1}$. A \emph{Chandra} observation by 
               \citet{yaqoob04} yields compatible results. 

\object{NGC 7172}: Intrinsic luminosity and column density estimates for this 
                source show a surprising range of variation. From 
                \emph{Beppo}SAX observations, \citet{dadina07} infers 
                $N_{\mathrm{H}} = 1.1 \times 10^{21}$ cm$^{-2}$
                (although this seems to be underestimated by a factor
                of 100 due to a typographical error) and 
                $F_{\mathrm{2-10 keV}} = 8.9 \times 10^{-12}$ erg s$^{-1}$ 
                cm$^{-2}$, yielding $L_{\mathrm{2-10 keV}} = 1.33 \times
                10^{42}$ erg s$^{-1}$. \citet{awaki06} on the other hand
                report $N_{\mathrm{H}} = (8.3 \pm 0.2) \times 10^{22}$ 
                cm$^{-2}$ and $L_{\mathrm{2-10 keV}} = 5.8 \times 10^{42}$ erg
                s$^{-1}$. \citet{risaliti02b} compiled observations of NGC 7172
                that had been executed between 1985 and 1997. Column densities 
                range between $(7  \sim 11) \times 10^{22}$ cm$^{-2}$ and 
                absorption corrected fluxes between $(0.9 \sim 7.7) \times 
                10^{-11}$ erg s$^{-1}$ cm$^{-2}$ with a weak trend toward lower
                fluxes with time. For our study we adopt the results of 
                Awaki et al. as they are intermediate ones. A high variability 
                of NGC 7172 in the hard X-ray band has been observed by 
                \citet{georga01} who report a flux variation by a factor of 
                $\sim 6$ within one week of observation. To account for this we
                set the uncertainty to 0.8 dex. 

\object{NGC 7213}: This object has been observed repeatedly by X-ray 
                satellites. The most recent results have been reported by 
                \citet{starling05} from XMM-\emph{Newton} observations 
                (Galactic $N_{\mathrm{H}} = 2.04 \times 10^{20}$ cm$^{-2}$, 
                $L_{\mathrm{2-10 keV}} = 1.7 \times 10^{42}$ erg s$^{-1}$),
                \citet{zhou05}, reanalysing archival XMM-\emph{Newton} data
                ($L_{\mathrm{2-10 keV}} = 1.68 \times 10^{42}$ erg s$^{-1}$ and
                \citet{bianchi04} who performed simultaneous observations with
                XMM-\emph{Newton} and \emph{Beppo}SAX ($N_{\mathrm{H}} = 2.04 
                \times 10^{20}$ cm$^{-2}$, $L_{\mathrm{2-10 keV}} = 1.72 
                \times 10^{42}$ erg s$^{-1}$). The agreement of these studies 
                is very good. We will use the arithmetic mean of these three 
                luminosity estimates.
    
\object{3C 445}: The three most recent observations of 3C 445 were all carried
                out with XMM-\emph{Newton}; \citet{shinozaki06} measured 
                $N_{\mathrm{H}} = 1.32^{+0.1}_{-0.2} \times 10^{23}$ 
                cm$^{-2}$ and -- in our Cosmology -- $L_{\mathrm{2-10 keV}} =
                2.6 \times 10^{44}$ erg s$^{-1}$. \citet{grandi07} find 
                $N_{\mathrm{H}} = 4^{+3}_{-2} \times 10^{23}$ cm$^{-2}$ and an
                unabsorbed flux of $F_{\mathrm{2-10 keV}} = 1.7 \times 
                10^{-11}$ erg s$^{-1}$ cm$^{-2}$ which translates to $1.2
                \times 10^{44}$ erg s$^{-1}$. The observations by 
                \citet{sambruna07} yield $N_{\mathrm{H}} = 
                2.25^{+0.62}_{-0.43} \times 10^{23}$ cm$^{-2}$ and, after 
                conversion to our Cosmology, $L_{\mathrm{2-10 keV}} = 8.1 
                \times 10^{43}$ erg s$^{-1}$. Given the complexity of the 
                source, these numbers agree very well. We therefore use the 
                mean values.

\object{NGC 7469}: From XMM-\emph{Newton} observations \citet{zhou05} derive a 
                luminosity of $L_{\mathrm{2-10 keV}} = 1.29 \times 10^{43}$
                erg s$^{-1}$. NGC 7469 was also observed with \emph{Chandra} 
                by \citet{jiang06} who find a purely Galactic absorption of 
                $N_{\mathrm{H}} = 4.9 \times 10^{20}$ cm$^{-2}$. After 
                correcting for the different Cosmology we use as well as for a 
                slightly different redshift, the luminosity they estimate, 
                translates to $L_{\mathrm{2-10 kev}} = 1.46 \times 10^{43}$ 
                erg s$^{-1}$. This is in good agreement with the result of 
                Zhou \& Wang. \citet{scott05} observed this object 
                simultaneously in the X-rays and UV domains with 
                \emph{Chandra}, FUSE and STIS. They do not state a 2-10 keV 
                flux in their paper. However, using the power law model shown 
                in their figure 1, we derive $F_{\mathrm{2-10 keV}} = 2.46 
                \times 10^{-11}$ erg s$^{-1}$ cm$^{-2}$. This, in turn, yields 
                $L_{\mathrm{2-10 keV}} = 1.45 \times 10^{43}$ erg s$^{-1}$. As 
                these three observations agree very well with each other, we 
                will use the mean luminosity and set the uncertainty to 
                0.2 dex.

\object{NGC 7674}: This object appears to be Compton-thick. \citet{malaguti98} 
                suggest an intrinsic luminosity of $L_{\mathrm{2-10 keV}} 
                \approx 10^{45}$ erg s$^{-1}$. For their estimate they assume
                the electron scattering material to have the same geometry as
                in the prototypical Seyfert 2 galaxy NGC 1068. Correcting for
                the different Cosmology they use, decreases this number to
                $\sim 4.4 \times 10^{44}$ erg s$^{-1}$.
                Another way to estimate the intrinsic 2-10 keV luminosity is
                via the correlation between the [O III]$_{\lambda 5007}$ and 
                2-10 keV fluxes that was found by \citet{panessa06}: 
                $\log F_{\mathrm{2-10 keV}} - \log F_{\mathrm{[O III]}}
                \approx 1.74$. \citet{dahari88} measured a flux of
                $F_{\mathrm{[O III]}} = 4.3 \times 10^{-13}$ erg s$^{-1}$ 
                cm$^{-2}$. \citet{bassani99} corrected this result for 
                intrinsic absorption and obtained
                $F_{\mathrm{[O III]}}^{\mathrm{int}} = (1.85 \pm 0.1) \times
                10^{-12}$ erg s$^{-1}$ cm$^{-2}$ with which, in turn, we obtain
                $L_{\mathrm{2-10 keV}} = 1.8 \times 10^{44}$ erg s$^{-1}$. 
                This result is in good agreement with the estimate by Malaguti
                et al., especially regarding the somewhat speculative nature
                of both methods. However, Bassani et al.'s absorption 
                correction of the [O III]$_{\lambda 5007}$ flux may suffer
                from the large slit ($2.7 \arcsec \times 4.0 \arcsec$)
                \citet{dahari88} used. If the Balmer decrement was affected by
                off-nuclear emission, the reddening could easily be 
                underestimated.
                In a private communication to \citet{malaguti98}, R. Maiolino
                reports $L_{\mathrm{[O III]}}^{\mathrm{int}} \approx 6 \times
                10^{44}$ erg s$^{-1}$. In the cosmology used by Malaguti et
                al., this corresponds to a [O III]$_{\lambda 5007}$ line flux
                of $1.5 \times 10^{-10}$ erg s$^{-1}$ cm$^{-2}$. This, in turn,
                yields $L_{\mathrm{2-10 keV}} \approx 1.5 \times 10^{46}$ erg 
                s$^{-1}$, i.e. a source that is two orders of magnitude
                brighter than estimated by Malaguti et al.
                For our analysis we will use the more conservative luminosity
                estimate of Malaguti et al. and Bassani et al. by using the 
                average of their estimates and allowing for an uncertainty of
                1 order of magnitude. An even higher luminosity, however, can 
                clearly not be ruled out. 
     
\object{NGC 7679}: The only available recent hard X-ray data of 
                this object seem to be the ones from \citet{dellac01} who use 
                ASCA and \emph{Beppo}SAX data. Their analysis yields
                $N_{\mathrm{H}} = 2.2^{+1.8}_{-1.4} \times 10^{20}$ cm$^{-2}$ 
                and $F_{\mathrm{2-10 keV}} = 5.6 \times 10^{-12}$ erg
                s$^{-1}$ cm$^{-2}$, the latter corresponding to 
                $L_{\mathrm{2-10 kev}} = 3.3 \times 10^{42}$ erg s$^{-1}$.

\end{appendix}

\end{document}